\documentclass[sigplan,10pt]{acmart}
\setcopyright{none}

\usepackage{xspace}
\usepackage{physics}
\usepackage{bm}
\usepackage{algorithmic}
\usepackage{algorithm}
\usepackage{verbatim}
\AtBeginDocument{%
  }

\begin{document}

\title{FastMPS: Revisit Data Parallel in Large-scale Matrix Product State Sampling}

\author{Yaojian Chen}
\email{yj-chen21@mails.tsinghua.edu.cn}
\affiliation{%
  \institution{Department of Computer Science and Technology, Tsinghua University}
  \city{Beijing}
  \country{China}
}

\author{Si-Qiu Gong}
\affiliation{%
  \institution{Hefei National Research Center for Physical Sciences at the Microscale and School of Physical Sciences, University of Science and Technology of China}
  \city{Hefei}
  \country{China}}

\author{Lin Gan}
\affiliation{%
  \institution{Department of Computer Science and Technology, Tsinghua University}
  \city{Beijing}
  \country{China}
}

\author{Yanfei Liu}
\affiliation{%
 \institution{National Supercomputing Center in Wuxi}
 \city{Wuxi}
 \country{China}}

\author{An Yang}
\affiliation{%
  \institution{Department of Computer Science and Technology, Tsinghua University}
  \city{Beijing}
  \country{China}}

\author{Yinuo Wang}
\affiliation{%
  \institution{Department of Computer Science and Technology, Tsinghua University}
  \city{Beijing}
  \country{China}}

\author{Chao-Yang Lu}
\affiliation{%
  \institution{Hefei National Research Center for Physical Sciences at the Microscale and School of Physical Sciences, University of Science and Technology of China}
  \city{Hefei}
  \country{China}}

\author{Guangwen Yang}
\affiliation{%
  \institution{Department of Computer Science and Technology, Tsinghua University}
  \city{Beijing}
  \country{China}}

\renewcommand{\shortauthors}{Trovato et al.}
\newcommand{\system}{\emph{Fast-MPS}\xspace}

\begin{abstract}
Matrix Product State (MPS) is a versatile tensor network representation widely applied in quantum physics, quantum chemistry, and machine learning, \emph{etc}. MPS sampling serves as a critical
fundamental operation in these fields. As the problems become more complex, the scale of MPS is rapidly increasing. Traditional data parallelism is limited by memory and heavy I/O in large-scale MPS. Model parallelism that can handle large-scale MPS imposes rigid process bindings and lacks scalability. This work proposes \system, a multi-level parallel framework for scalable MPS sampling. Our design combines data parallelism across samples with tensor parallelism along bond dimensions. We eliminate memory and I/O pressure through compression and overlapping, and revive data parallel in large-scale MPS sampling. We evaluate our approach on Gaussian Boson Sampling, a representative and demanding application. \system achieves over $10\times$ speedup compared to existing simulators, scales to thousands of processes, and enables simulations with 8,176 sites and bond dimension $\chi = 10^4$, significantly outperforming the state of the art. 
\system has demonstrated great potential in high-performance tensor network applications.
\end{abstract}



\maketitle
 
\section{Introduction}

\emph{Matrix Product States} (MPS)\cite{perez2006matrix} have emerged as one of the most versatile tensor network representations for modeling high-dimensional quantum systems. Originally developed in condensed matter physics to efficiently approximate weakly entangled states, MPS have since become central to areas such as quantum chemistry, quantum circuit simulation, and quantum machine learning~\cite{schollwock2011density, chan2011density, orus2014practical, stoudenmire2016supervised}. The advantage of MPS lies in their ability to compress exponentially large state spaces into a sequence of low-rank tensors, where the bond dimension $\chi$ controls accuracy. \emph{Sampling from an MPS} is a critical primitive across disciplines. In quantum many-body physics, it enables measurement emulation and Monte Carlo estimation~\cite{orus2014practical, verstraete2008matrix}. In machine learning, MPS form the backbone of generative models known as \emph{Born machines}, which require efficient sampling to learn and generate high-dimensional distributions~\cite{han2018unsupervised, glasser2019expressive}. In the design of quantum-computing hardware, MPS sampling underpins classical simulation of photonic quantum computing experiments, including Gaussian Boson Sampling (GBS)~\cite{zhong2020quantum, zhong2021phase, deng2023gaussian, madsen2022quantum, liu2025robust}. 

In recent years, computational physicists began to pay more attention to large-scale, strongly correlated systems, which require MPS to have higher bond dimensions $\chi \sim 10^4$, larger sites $M \sim 8000$, and more samples $N\sim 10^7$. With the $NM\chi^2$ growth time complexity, MPS sampling should be redesigned to adapt high-performance computing clusters, with efficient parallel algorithms. However, existing approaches remain limited in scalability and flexibility, as MPS sampling is fundamentally sequential: each site measurement depends on the contraction outcome of preceding sites. The strong data dependency hinders both model parallelization and tensor parallelization. Moreover, $M$ tensors, with size $\chi^2$ (GB-level per tensor if $\chi\sim 10^4$), bring great storage and I/O pressure. The maximum scale of the data parallel scheme\cite{shang2022large} is $\chi\sim1024, M\sim200$. When the MPS system exceeds memory capacity, repeatedly
loading tensors for each batch severely limits efficiency. The parallel scheme that assigns one process per site\cite{oh2024classical} is able to extend $\chi$ to $10^4$, ensuring sufficient accuracy, but suffers from rigid process binding, startup imbalance, and heavy communication. To keep pace with the rapid advances in quantum experiments\cite{deng2023gaussian, liu2025robust}, where thousands of sites and $10^4$ bond dimension are becoming standard, these limitations must be addressed to fully harness MPS sampling.

In this work, we revisit MPS sampling from a data-parallel and memory-centric perspective. The potential of data parallelism is reinvigorated by compressing memory and masking I/O. We combines data parallelism across independent samples with tensor parallelism along bond dimensions, overcoming the rigid scaling barrier. Our design incorporates three key innovations: (1) a scalable parallel batch sampling algorithm that balances computation, communication, and memory locality; (2) an adaptive mixed-precision strategy that preserves numerical stability while exploiting tensor-core acceleration; and (3) customized kernels and dynamic bond dimension allocation to reduce complexity. These optimizations break the scaling bottleneck of prior methods, enabling efficient simulation at unprecedented scales.

We demonstrate the effectiveness of our approach on Gaussian Boson Sampling, a demanding and representative application for MPS sampling. Our system, \system, achieves more than $10\times$ performance improvement over prior MPS-based simulators, scales to thousands of processes, and successfully handles simulations with 8,176 sites and bond dimension $\chi = 10^4$---well beyond the previous state of the art. We believe our work establishes parallel MPS sampling as a critical building block for both quantum simulation and broader tensor network applications.

\section{Background}

\subsection{Benchmark: Gaussian Boson Sampling}
Gaussian Boson Sampling (GBS)\cite{hamilton2017gaussian} has been considered a promising approach to show quantum advantage via photons. In GBS experiments, input photons go through a beam splitter network, and are detected at the output modes. The exact classical simulation methods, including \emph{Torontonian}\cite{PhysRevA.98.062322, li2021benchmarking}, \emph{Hafnian}\cite{quesada2022quadratic, bulmer2022boundary}, suffer from exponential scaling with photon numbers. However, inevitable photon loss and noise in actual experiments, undermine the quantum advantage of existing photonic quantum computers\cite{gao2018efficient, garcia2019simulating, oh2023classical}, as noise does not contribute to the quantum computing power. A recent approach\cite{oh2021classical} introduced MPS as a more efficient representation of the GBS final state. In the following work\cite{oh2024classical}, the output state of a noisy GBS is divided into a smaller MPS and a classical noise term. The quantum part consists of fewer effective photons, indicating that one can achieve sufficient accuracy with a small bond dimension $\chi$. Therefore, MPS-based method\cite{oh2024classical} shows the ability to simulate state-of-the-art GBS experiments\cite{madsen2022quantum, deng2023gaussian} up to 288 sites in 1 hour, using hundreds of GPUs. MPS sampling dominants the time cost in practical simulation. At the same time, since GBS data is standardized and easy to obtain, it is suitable as a benchmark. Therefore, we test the effect of MPS sampling optimization on GBS data.

\subsection{Parallel MPS Sampling}

\begin{figure}[t]
\centerline{\includegraphics[width=0.45\textwidth]{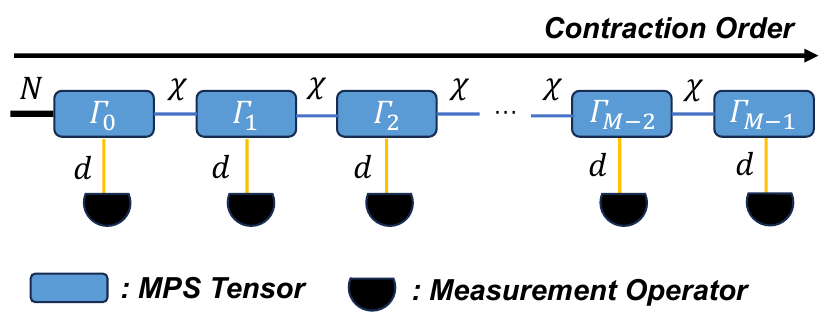}}
\caption{MPS sampling workflow. The open index $N$ denotes to $N$ samples. Contraction is sequentially performed from left to right. Just after the i-th contraction, measurement happens to drop the physical index $d$.}
\label{mps}
\end{figure}

MPS has long been used to represent weakly entangled quantum states\cite{schollwock2011density} in quantum computing and quantum many body problems, due to its representation power and polynomial memory requirement. A typical MPS, as Figure~\ref{mps} shown, retains the physical indices $d$ of the state vector and decomposes the state vector into $M$ tensors linked by auxiliary indices $\chi$. Compared with the original memory cost $O(d^M)$, MPS consists of $M$ tensors and requires only $O(Md\chi^2)$ space. Bond dimension $\chi$ comes from the truncation of higher-order entanglement. Simulating highly entangled quantum systems often requires large $\chi$ for accuracy. 

The canonical MPS sampling workflow follows a sequential procedure, as demonstrated in Figure~\ref{mps} (To facilitate understanding, we mark the shape of each tensor involved below): (1) The left boundary tensor $\Gamma_0$ is first measured as the left environment ($N, \chi$). (2) Starting from $\Gamma_1$, at the $i-th$ site, the MPS tensor $\Gamma_i, (\chi, \chi, d)$ is contracted into the left environment ($N, \chi$). (3) Before proceeding to the next site, the left environment ($N, \chi, d$) is measured and drops the physical dimension $d$. Then shape of left environment goes back to ($N, \chi$). While exact and stable, this process is inherently sequential, and its complexity grows as $O(NM \chi^2 d)$. 

For small $\chi$ (e.g. $< 2000$) and $M$ ($\sim 100$), the entire MPS can be stored in the global memory of a single GPU or a non-uniform memory access (NUMA) node. In this case, data parallelization
across samples\cite{shang2022large} is straightforward. When the MPS size exceeds a single process’s memory, data parallel schemes repeatedly swap tensors between memory and disk, severely impacting performance. As a result, naive data parallelization becomes I/O-bound.

\begin{figure}[t]
\centerline{\includegraphics[width=0.45\textwidth]{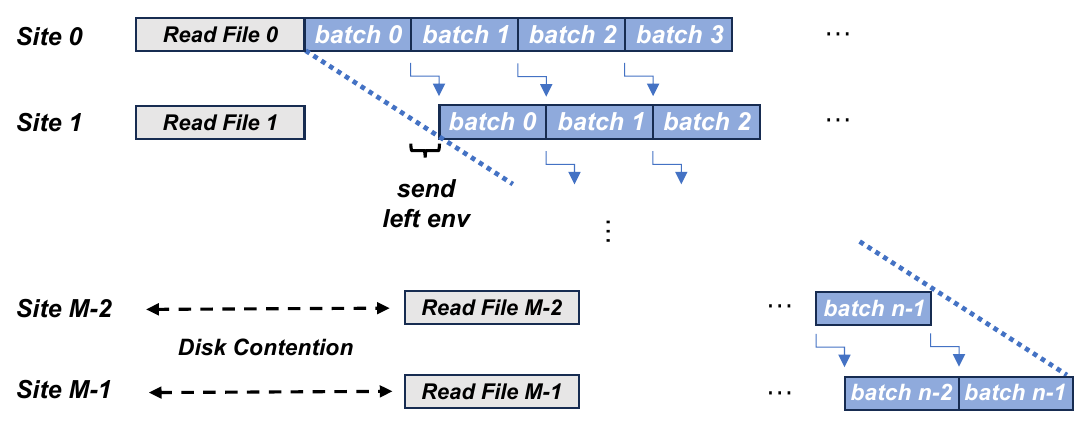}}
\caption{Distributed sampling in \cite{oh2024classical}. Each process handles one site and stores one MPS tensor. Every sample is calculated by all processes collaboratively. When a macro batch is finished at process $i$, the left environment pre\_tensor will be non-blocking sent to process $i+1$. Disk contention may occur due to the massive I/O at the beginning.}
\label{distribute-mps}
\end{figure}

A potential solution is to do model parallelization, which distributes the whole MPS into multiple devices. A recent work\cite{oh2024classical} fixed the number of processes to $M$ and let each process hold an MPS tensor. Although scalability is sacrificed, this approach can handle massive sampling with $\chi\sim10^4, N\sim10^7$. They carefully used computation-communication overlapping to overcome sequential data dependencies, as Figure~\ref{distribute-mps} shown. This scheme fully utilizes the locality of $\Gamma$, as each $\Gamma_i$ is loaded only once. The left environment is sent point-to-point to the successor process. The large batch ($N\sim10^7$) is divided into intermediate-size macro batches ($N_1 \sim 10^3-10^4, N=n_1N_1$), to conserve memory and establish a pipeline. Once process $i$ finishes a macro batch, the result will be sent in a non-blocking manner to process $i+1$, and process $i$ will continue to deal with the next macro batch. Therefore, the performance model of this scheme is:
\begin{equation}
    T_{all} = T_{0, Read} + n_1\max_iT_{i, N_1} + \sum_{i=0}^{M-1} (T_{i, N_1} +T_{i, comm})
    \label{model}
\end{equation}

$T_{i, N_1}$ denotes the time cost of a $N_1-$sample macro batch calculation, and $T_{0, Read}$ is the I/O time at the 0-th site. Here we use a general $T_{i, N_1}$ and $T_{i, comm}$, as the time cost in different sites may vary. 

There are many critical problems in this scheme. (1) The most fatal one, the fixed number of processes not only hinders scaling, but also eliminates the opportunity for users with limited computing resources to perform simulations. (2) As massive I/O is concentrated at the beginning, disk contention may further enlarge the startup cost due to the limitation of I/O bandwidth. (3) The initialization of pipeline is expensive. With thousands of sites, the last process needs to wait for $M-1$ times calculation and communication. (4) The whole communication amount is $O(NM\chi)$. At each step, the computation-to-communication ratio (CCR) is $\frac{s\chi^2d}{s\chi}=\chi d$. With complex64 precision, $d\sim3$, the exact CCR is near 3700 FLOPs/byte, which is easy to fall to communication-bound (CCR threshold is 3120 FLOPs/byte for A100 GPU with Infiniband). To address these issues, a new parallel scheme is required.


\section{Innovations}
\subsection{Performance Model of Data Parallel Sampling }
\label{parallel}

\begin{figure}[t]
\centerline{\includegraphics[width=0.4\textwidth]{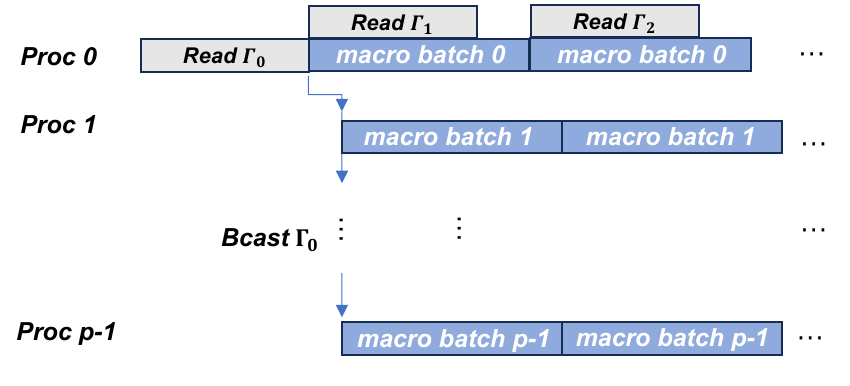}}
\caption{Data parallelization scheme. Process 0 takes all I/O and bcasts the MPS tensor $\Gamma$ to all processes. All processes need to go through all sites and calculate independent macro batches. I/O and calculation could be overlapped. With $n_1$ macro batches in total, this workflow will repeat $n_1/p$ times.}
\label{dataparallel}
\end{figure}

\begin{figure*}[t]
\centerline{\includegraphics[width=0.95\textwidth]{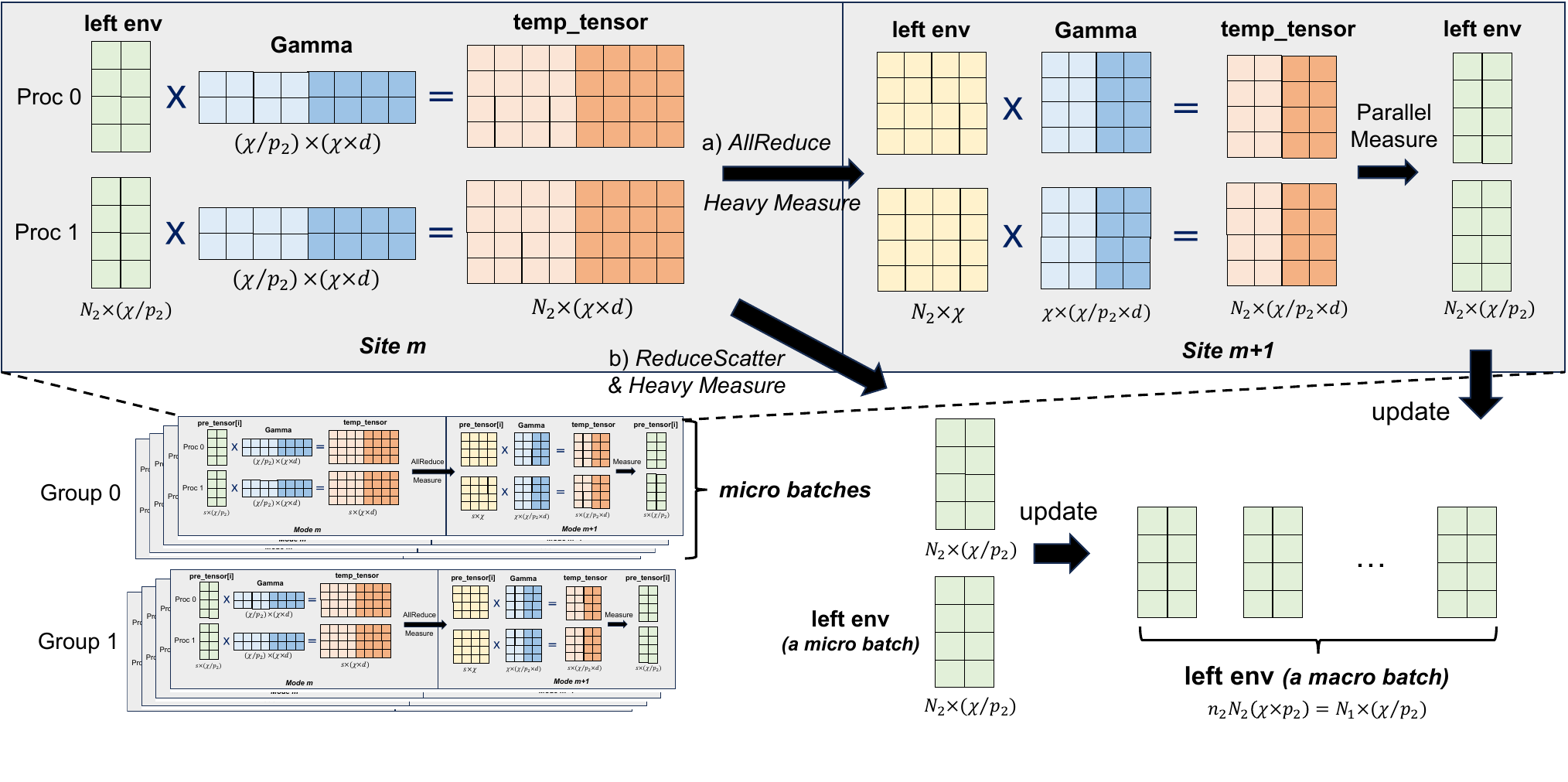}}
\caption{The multi-level parallel scheme. At the task-level, independent samples could be distributed to groups. In each group, we do tensor parallelization on $\Gamma$ for large $\chi$. There are two version of method: \textbf{a) Double-Site division} and \textbf{b) Single-Site division}, using AllReduce and ReduceScatter for communication, respectively.}
\label{allreduce}
\end{figure*}

The proposed parallel scheme is designed to preserve the advantages of the previous method, such as computation-communication overlap and minimum I/O operations, while avoiding its disadvantages, including process binding, startup costs, and load balancing issues. A natural solution is data parallelization, since $N\sim10^7$ independent samples offer enough parallelism. Figure~\ref{dataparallel} illustrates the data parallelism scheme. Each process holds $N_p = N / p$ samples, organized as $n_1/p$ macro batches. Process 0 is responsible for loading and broadcasting $\Gamma$. By employing a double-buffer and an additional thread on the CPU, overlap between computation and I/O, as well as between computation and communication, is achieved. The ideal performance model of this scheme is:
\begin{equation}
    T_{all} = T_{0, Read} + T_{0, comm} + \frac{n_1}{p}\sum_{i=0}^{M-1}T_{i, N_1}
\end{equation} 
Compared with the theoretical time cost of model parallel ($p=M$ to ensure same resource), the difference is $\sum_{i=1}^{M-1} (T_{i, N_1} +T_{i, comm}) + T_{0, N_1} + N(\max(T_{i, N_1}) - mean(T_{i, N_1})) > 0$. This indicates that, theoretically, data parallelism outperforms the model parallelism approach. The saved time comes from canceling expensive pipeline initialization and resolving load imbalance. However, this ideal model deviates significantly from practical performance. Achieving the ideal performance model necessitates meticulous memory management.

The full overlap between I/O and computation relies on the assumption that $T_{comp} > T_{IO}$. With macro batch size $N_1$, the computation-I/O ratio at one site is $N_1\chi^2d/(\chi^2d)=N_1$. Therefore, insufficient memory for large macro batches leads to I/O dominating overall runtime. Considering that the common NVMe SSD reading bandwidth is $\sim 5GB/s$, and the $156TFLOPS$ peak performance of A100 GPU, a safe $N_1$ should be $\sim 10^5-10^6$. For CPU, with lower computation power, $N_1$ could be much smaller to enable larger parallelism. Naively, the allocated memory consists of an $\chi^2d$ MPS tensor $\Gamma_i$, the intermediate result $N_1\chi d$, and the left environment $N_1\chi$. We further divided the macro batch into $n_2$ micro batches with $N_2$ samples each to eliminate the memory of intermediate result, as $N_1 \gg N_2d$. Therefore, the memory demand will be (complex double precision):
\begin{equation}
Mem = Mem_{left\_env} + Mem_{\Gamma} = (N_1\chi d + \chi^2d) \times 16B
\label{mem}
\end{equation}

Therefore, enlarge $N_1$ under a fixed memory capacity to reach the threshold is the core of our following approach. Tensor parallel and low-precision will be designed to reduce memory requirement.

\subsection{Single- and Double-Site Tensor Parallelization}

As $\Gamma$ grows quadratically with $\chi$, it will rapidly fill the memory with large $\chi$. For $\chi = 20,000$ and $d = 3$, a single $\Gamma$ requires 19.2 GB (FP64), exceeding the global memory of many GPUs and AI chips. Although some data center GPUs can handle this tensor, the remaining memory supports only $N_1\sim 10^4$, which leads to I/O-domination.

To address these issues, we further proposed a tensor parallel framework to split $\Gamma$ from $\chi$ axis, as illustrated in Figure~\ref{allreduce}. $p=p_1 \times p_2$ processes are organized into $p_1$ communication groups. Each group handles $N_p = N/p_1$ samples, packed by $n_1 / p_1$ macro batches. In a certain group (including $p_2$ processes), initially, $\Gamma_0$ is sliced along its only axis of dimension $\chi$ and loaded onto $p_2$ processes in parallel. Then, it is converted as left environment and measured, with shape $(N_1, \chi/p_2)$. The following MPS tensors, $\Gamma_i$, will be sliced along its first axis. Then, $p_2$ processes cooperate to perform a split-K GEMM. The temporary result is then measured to reduce its physical dimension. Before next site, the temporary tensors on $p_2$ processes should be (1) reduced to sum up the split-K results and form the left environment, (2) sliced and distributed along $\chi$ axis to prepare the next split-K GEMM with the next $\Gamma$. Since each process only needs its own sliced sub-tensor to be reduced, ReduceScatter can simultaneously achieve these two goals, as Figure~\ref{allreduce} b) shown. In detail, split-K GEMM and communication is performed in micro batch. The generated left environment should be stored in global memory until the entire macro batch is updated. This scheme is called single-site tensor parallelization, suitable for bandwidth-dominated communication. Due to frequent collective communication, Single-site scheme is not friendly for high-latency communication like PCIE. Moreover, to reduce the communication volume by a factor $d$, measurement is performed before communication. That means one should do non-distributed measurement, which leads to $p_2$ times overhead.

For latency-dominated condition, we designed double-site tensor parallelization.  The double-site scheme cut the number of communications by half  to reduce latency, as Figure~\ref{allreduce} a) shown. In this scheme, two neighbor sites are processed together. The handling of odd sites is the same as the single-site method. The key difference is that ReduceScatter is replaced by AllReduce. AllReduce distributes the entire left environment to all $p_2$ processes. Then, $\Gamma$s in even sites are split into segments of shape $\chi \times (\chi/p_2 \times d)$. At this time, the distribution of generated left environment in an even site just matches the input of an odd site, which means this GEMM is locally performed without communication. As the unmeasured left environment is distributed, measurement can be naturally parallel.

The choice of single-site or double-site scheme is based on specific hardware, network topology and the communication algorithms. The two schemes have the same average communication volume and similar GEMM performance at one site. Considering data dependence, these collective communication can not be properly overlapped. So an intuitive performance model of site $i$ can be expressed as: 
\begin{equation}
    T_{i, N_2} = T_{GEMM} + T_{Measure} + \frac{Comm Volume}{Bandwidth}
\end{equation}
Therefore, the difference in performance comes from the bandwidth, latency, and measurement performance. If measurement accounts for a high proportion, double-site scheme will be better.

The advantage of tensor parallelism is two-fold. (1) Two-level process organization enables full utilization of high bandwidth within nodes. Massive collective communication between tensor slices is handled by fast hardware (like NV-Link). (2) Splitting tensors reduces memory requirements. Compared with pure data parallelization, MPS tensor $\Gamma$ is also distributed in $p_2$ processes. This enables larger $\chi$ to achieve higher accuracy, and allows larger macro batch for computation-I/O overlap. 

\subsection{Adaptive Mixed Precision}
\subsubsection{Numeric Stability Maintaining}

\begin{figure}[t]
\centerline{\includegraphics[width=0.45\textwidth]{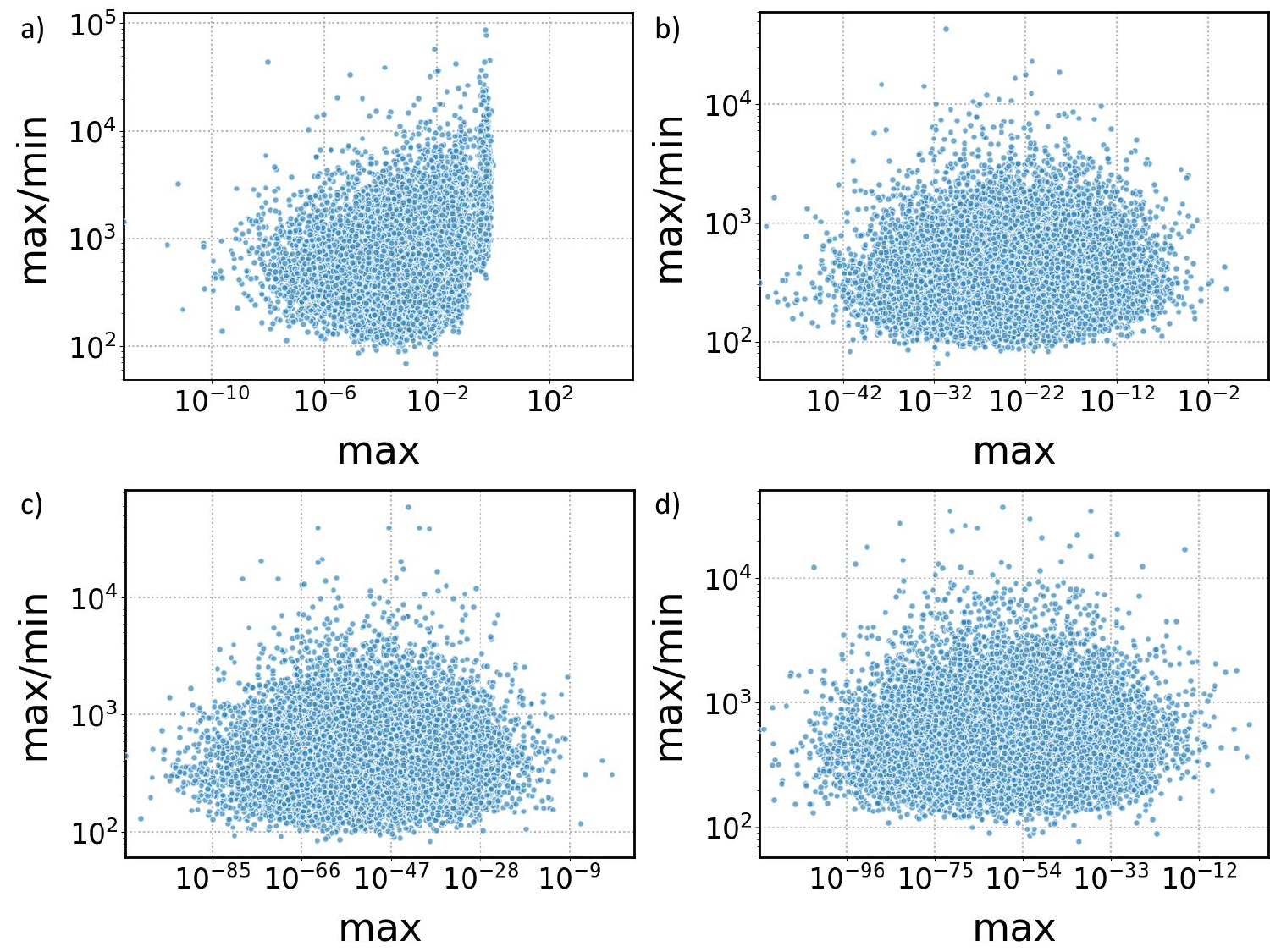}}
\caption{Visualization of the data distribution in left\_env $(N_2, \chi)$ with the variation of site. Each data point represents a sample. The horizontal axis shows the max value of each sample (i.e. $max(left\_env, axis=1)$). The vertical axis shows ratio of the maximum value to the minimum value of each sample. a) site = 450. b) site = 2000. c) site = 5000. d) site = 7150.}
\label{range}
\end{figure}

\begin{figure}[t]
\centerline{\includegraphics[width=0.38\textwidth]{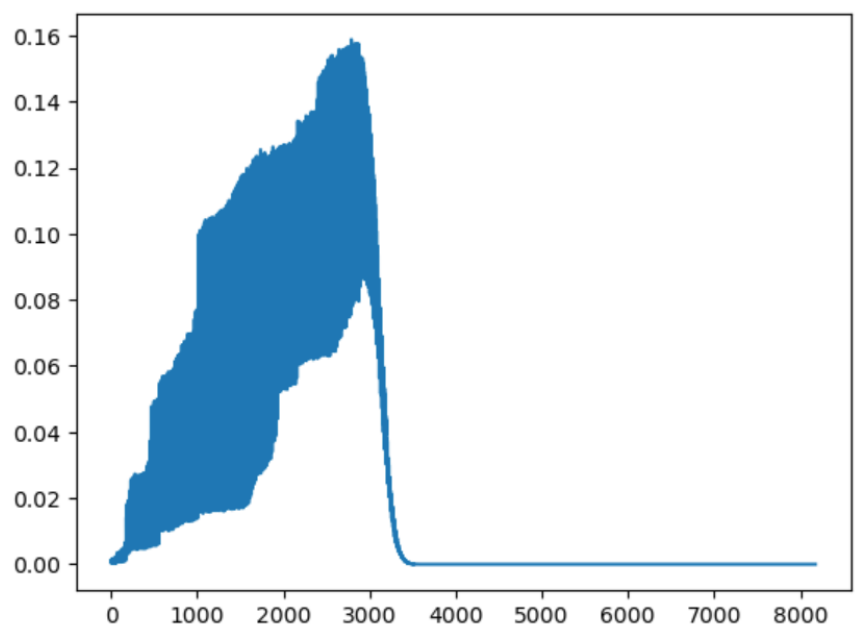}}
\caption{Sampling failed at a large site due to underflow. At site-3000 left environment becomes a 0-tensor. Testing on a 8176-site simulation data with 16.54 actual squeezed photons. The horizontal axis is the sites. The vertical axis is the average number of photons at a site.}
\label{decay}
\end{figure}

Using mixed precision in MPS is promising, since the hotspot is tensor operations. The numerical stability and floating-point truncation error of matrix multiplications have been discussed in previous works\cite{higham2002accuracy, higham2022mixed}. Tensor contraction based on matrix multiplication could be performed by tensor cores, whose block FMA summation scheme will further reduce error accumulation. In the existed implementation\cite{oh2024classical}, there is an auto-scaling method to avoid underflow. However, despite these efforts, the experimental results, as shown in Figure~\ref{decay}, still indicate the need for high precision up to FP64. It is a tough decision to switch to FP64 or higher precision, as it would cause a significant performance downgrade. On a single A100 GPU\cite{choquette2021nvidia}, the peak performance of FP64 is 9.5TFLOPS, while that of TF32 is 156TFLOPS. The huge performance gap makes it worthwhile to look for opportunities to maintain low precision.

The high-precision demands comes from the extremely wide numerical range. The magnitudes of the intermediate results (left environment at each site) generally decrease after contractions. $\Gamma_i$ of each site have similar effects on reducing the magnitude by approximately $k$. Thus, the mean value of left environment $\mu$ in a certain site $i$ should follow: 
\begin{equation}
    \mu_{i} \sim \mu_010^{-ik}
\end{equation}

As $k$ is an uncertain number that varies with the site and data set, fixed scaling factors commonly fail to find a balance between underflow and overflow. The previous implementation\cite{oh2024classical} designed an auto-scaling method to mitigate the problem. They used the maximum value of the intermediate tensor as the dynamic scaling factor to shift the data range towards 1. However, this scaling method can only address the data shift from 1, but not the data range expansion. From Figure~\ref{range} a) to d), with increasing site, though the maximum data point is still $<1$, the range of data expands rapidly. Even without considering the differences within the samples, the maximum values (horizontal axis) of different samples differ by hundreds of orders of magnitude. This results in Figure~\ref{decay}. With large sites, data range expansion sufficiently causes underflow, and rapidly infects the entire left environment, making the subsequent results 0. Thus, the scaling method should take on the task of narrowing the range. 

\begin{algorithm}
	\caption{Measurement}
	\label{alg2}
	\renewcommand{\algorithmicrequire}{\textbf{Input:}}
	\renewcommand{\algorithmicensure}{\textbf{Output:}}
	\begin{algorithmic}
		\REQUIRE Unmeasured Left Environment $\mathbf{temp\_tensor}$, Coefficient Vector $\Lambda$
		\STATE \texttt{/* Get probs: $(N_2,\chi, d), (\chi) \rightarrow (N_2, d)$ */}
            \STATE $\mathbf{probs = dot(temp\_tensor, \Lambda)}$
            \STATE \texttt{/* Normalization: $(N_2, d) \rightarrow (N_2, d)$ */}
            \STATE $\mathbf{probs = cumsum(probs / sum(probs, axis=1), axis=1)}$
            \STATE \texttt{/* Initalize Random Threshold $(N_2, d)$*/}
            \STATE $\mathbf{threshold = repeat(rand(N_2), d)}$
            \STATE \texttt{/* Collapsed States in $N_2$ Samples */}
            \STATE $\mathbf{samples = sum(threshold > probs, axis=1)}$
            \STATE \texttt{/* Update Left Env $(N_2,\chi, d), (\chi) \rightarrow (N_2, \chi)$*/}
            \STATE $\mathbf{Left\_Env = temp\_tensor[range(N_2), :, : samples]}$
		\ENSURE $\mathbf{samples, Left\_Env}$
	\end{algorithmic}
\end{algorithm}

The opportunity comes from the measurement algorithm (Alg.~\ref{alg2}) and the data distribution in Figure~\ref{range}. In this algorithm, the probability and the final photon number are linearly derived from temp\_tensor. There is no data exchange among the samples. Figure~\ref{range} further illustrates that the extreme magnitude differences occur between samples. Within a sample, the maximum data range is $10^6$, which is well above the underflow threshold. Therefore, we can use a custom scaling factor for each sample, typically the sample-wise maximum value. The normalization further cancels the restoration after scaling and eliminates the need to maintain a reverse scaling vector. As a result, we can use TF32 in our calculations up to thousands of sites, thereby unleashing the powerful computing performance of tensor cores. We are currently cautious about fully transitioning to FP16/BF16 calculations. There are primarily two reasons. (1) ComplexHalf is not well supported in existing libraries (CuPy 13.3.0, Torch 2.5.1). (2) The maximum data range in a sample is near $10^6$, which exceeds the representation ability of the valid bits of FP16. This implies there may be more rounding and truncation error. The experimental FP16 version is developed only for datasets with $M < 500$.

\subsubsection{Low-precision Storage}
TF32 precision provides a leap in computing power. However, to improve overall performance, data throughput should keep pace with the computation. As discussed above, memory issues hinder performance in various ways, including I/O, host-device memcpy, communication, and pipeline overlap. These operations mainly consist of data movements, which are insensitive to errors and bandwidth-limited. Further, according to theoretical results and our tests, truncated error of inputs is controllable. Thus, using low precision to store and transfer data is advantageous. 

In our design, MPS tensors $\Gamma$ are pre-calculated and stored as FP16, and it will maintain the precision until contraction. As a result, FP16 $\Gamma$ reduces the overhead of I/O, host-device memcpy and bcast to half. The reduction of these overheads greatly relieves the pressure on overlapping, since a smaller macro batch $N_1$ is sufficient achieve the overlapping threshold. It worth noticing that $\Gamma$ is converted to TF32 before calculation. Therefore, FP16 $\Gamma$ does not reduce memory usage. At the same time, FP16 storage of the left environment provides the opportunity for doubling $N_1$ with the same memory cost.

The overlapping relationship with fp16 storage can be expressed as 
$2T_{i, N_1} > \tfrac{1}{2}T_{i, IO}$, releasing the pressure of overlapping I/O by one-fourth. Increasing the macro batch size by a factor of two reduces the total number of macro batches by half. Under the same computational resources, this directly halves the number of iterations in the data-parallel workflow. Considering that the number of I/O operations is reduced by half and the speed of each I/O operation is doubled, the total I/O load of the file system is reduced to 1/4.


\subsection{MPS-GBS Customized Optimization}
\subsubsection{Matrix Exponential For Displace Operator}
In GBS simulation, displacement calculations constitute a significant portion of the overhead, especially for a small bond dimension $\chi$. The time cost in sampling is mainly due to the matrix exponential (expm). The calculation of expm itself is complicated, and the general implementation in Eigen\cite{eigenweb} and SciPy cannot be directly extended to GPUs. Although with large $\chi$, the complexity of displacement, i.e. $Nd^3$, is ignorable, the low efficiency and non-GPU execution will close the gap, and make displacement take $>50\%$ time according to our profiling. Here we utilize the structured sparsity to simplify the calculation and accelerate the displacement on GPU.

The random displacement involves the displacement operator $e^{\mu a^{\dag}-\mu^{*}a}$, where $a,a^{\dag}$ are creation and annihilation operators, $\mu$ is a random complex number\cite{radmore1997methods}.  

\begin{figure}[t]
\centerline{\includegraphics[width=0.40\textwidth]{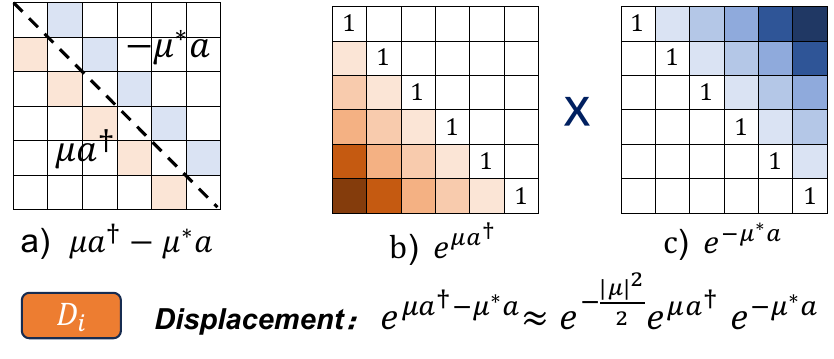}}
\caption{The decomposition and approximation method for displacement operator}
\label{expm}
\end{figure}

Figure~\ref{expm} shows the decomposition and customized approximation method for the displacement operator. a) shows that the matrix $\mu a^{\dag}-\mu^{*}a$ is $n \times n$ tridiagonal with zero diagonal elements. b) and c) indicate that $e^{\mu a^{\dag}}$ and $e^{-\mu^{*}a}$ are straightforward to compute. Thus, decomposing the exponent into two parts simplifies the calculation. In infinite-dimensional Hilbert space, the commutation relation $[a,a^{\dag}]=1$ is held. This can be utilized to perform the approximation on Zassenhaus formula\cite{casas2012efficient} in the $n$-dimensional case: 
\begin{equation}
    e^{\mu a^{\dag}-\mu^{*}a}\approx e^{-\frac{\abs{\mu}^2}{2}} e^{\mu a^{\dag}}e^{-\mu^{*}a} \label{con:zass approximation}
\end{equation}
We designed random test to evaluate the accuracy. If error is not ignorable, we will add a correction term with a diagonal matrix $e^{[\mu a^{\dag}, \mu^{*}a]}$. This will only introduce an extra GEMV with size $<10$. The approximation method in (\ref{con:zass approximation}) preserves the sampling results without significant loss. Moreover, the computational complexity is reduced to multiplying a lower triangular matrix by an upper triangular matrix, which reduces the computational time by a factor of more than 10. 

Each sample has its own $\mu$, so in practice we organize this process in a batched manner. If we simply let each thread calculate a matrix exponential, there will be significant bank conflicts on GPUs, as each thread should be responsible for contiguous memory sections. Noticing that both $e^A$ and $e^B$ are generated element-wise, we transpose the batch dimension to the last position. Then the memory that threads in the same warp are responsible for is interleaved. 

\begin{figure}[t]
\centerline{\includegraphics[width=0.4\textwidth]{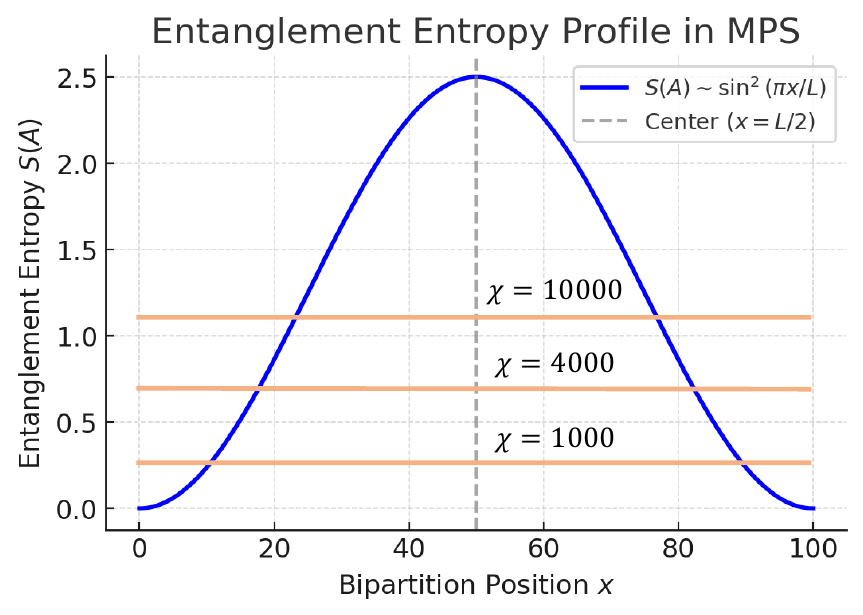}}
\caption{The Blue curve shows the entanglement distribution in MPS, quantified by entropy. The orange lines demonstrates the complexity using a certain bond dimension.}
\label{entangle}
\end{figure}

\subsubsection{Dynamic Bond Dimensions}
Bond dimension indicates the strength of entanglement\cite{perez2006matrix}. According to the area law for the entanglement entropy \cite{Eisert_2010}, entanglement increases gradually from the edges to the center. As a result, the bond dimension should be non-uniformly distributed in the MPS. In the context of exact simulation, the closer to the center, the higher the bond dimension. As Figure~\ref{entangle} shows, a fixed bond dimension is always redundant for the edges. Thus, a dynamic bond dimension can be set for each site, based on the entanglement. Only the region enclosed by the bond dimension line and the entanglement curve needs to be calculated. 

Specifically, in MPS-GBS, the same rule applies. The entanglement is identified in an $n_p^M$-dim Hilbert space, where $n_p$ denotes the thermal photon number and $M$ is the number of sites. The size of the Hilbert space is the theoretical upper bound of the bond dimension. In \cite{oh2024classical}, an error filter is designed to select $\chi$ high-amplitude points. In Figure~\ref{entangle}, the truncation error of a fixed bond dimension is mostly due to the central sites (the region above the bond line). Thus, we modified the filter to be more aggressive at the edges to reserve fewer points, while keeping the truncation error under control.

\section{Evaluation}
\subsection{Validation}
\begin{figure}[t]
\centerline{\includegraphics[width=0.45\textwidth]{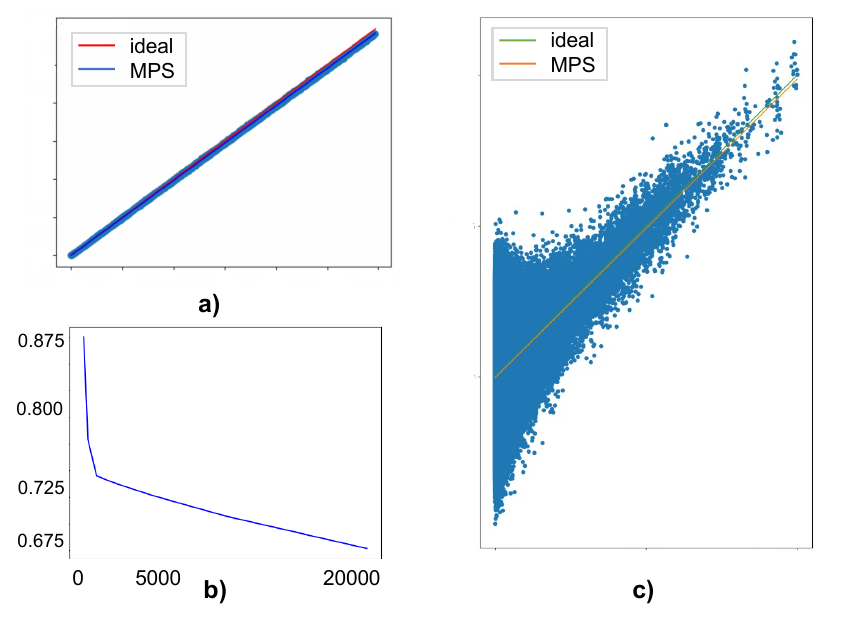}}
\caption{Validation results of MPS-GBS simulation on a 8176-site simulating data with 16.54 actual squeezed photons. a) First-order correlation. Slope: 0.97(simulation) $\sim$ 1 (ideal) b) Maximum truncation error among all sites with $\chi$. c) Second-order correlation. Slope: 0.96(simulation) $\sim$ 1 (ideal). Results in a) and c) uses parameters: $d=3, \chi=10000, 10M$ samples in total.}
\label{validation}
\end{figure}

In the simulation of Jiuzhang and Borealis, we keep the same parameters with \cite{oh2024classical}, and obtained strictly consistent sampling results using the same random seeds. For the large-scale simulations, we applied correlation functions, which are crucial evidences of claiming quantum advantage, for validation. Aligned with \cite{oh2024classical}, we evaluated the 1-st and 2-nd correlations (Figure~\ref{validation} a), c)) which covered the primary error contribution. Our results are basically consistent with the ideal slope. That means we expand the simulation scale from $M\sim288$ to $M=8176$, with $\chi=10000$ and 16.54 actual squeezed photons, well beyond the state-of-the-art. We also evaluated the error of the approximation for expm. The relative error at the elements which we care about, are less than $0.2\%$.

\begin{figure*}[t]
\centerline{\includegraphics[width=0.95\textwidth]{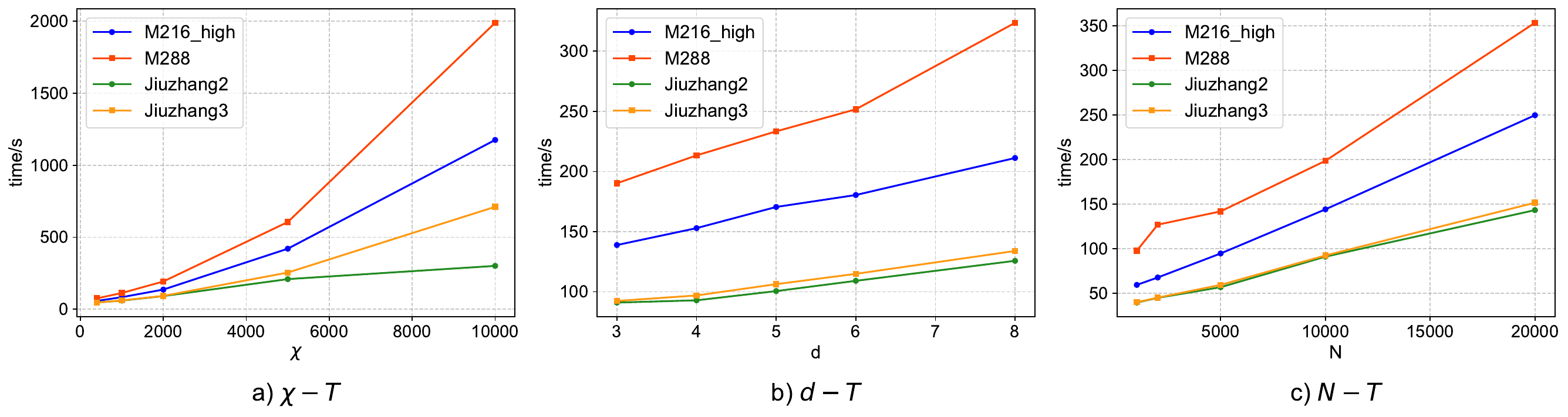}}
\caption{Time cost with bond dimension $\chi$, physical dimension $d$ and micro batch $N$ on state-of-the-arts experiments\cite{zhong2021phase, madsen2022quantum, deng2023gaussian}. All results are evaluated on a single A100 GPU. 20 pieces are performed to ensure overlapping. a) Variation of time cost with $\chi$. $d$ and $N$ are fixed as 3 and 20000. b) Variation of time cost with $d$. $\chi$ and $N$ are fixed as 2000 and 20000. c) Variation of time cost with $N$. $\chi$ and $d$ are fixed as 2000 and 3.}
\label{parameter}
\end{figure*}

The results are generally in line with expectations, but we still need to discuss the truncated error in depth because we applied a smaller bond dimension $\chi$ in large-scale experiments, which may bring potential error risks. As a key approximation in MPS, truncated $\chi$ accounts for the main source of error. Figure~\ref{validation} b) shows a decay of the truncation error at the middle sites with increasing bond dimension $\chi$ in the 8176-site simulation. Even with $\chi=20000$, the maximum truncated error is $\sim 0.675$. This requires some analysis to justify the results.

Here we propose a semi-quantitative method to evaluate the impact of truncation error on first order correlation. The ideal first order marginal distribution is $p = 1- \eta p_0$, where $\eta$ is the probability of the initial 0-th state, and $p_0$ denotes to the probability of being moved to the 0-th state by displacements. With truncation error, the first order distribution of MPS is $p' = 1 - \eta'p_0$. The slope is defined as $k = p'/p = 1 - \frac{\eta - \eta'}{\eta - 1/p_0}$. The truncation error $\epsilon$ comes from the discarded data points, accounting for the only difference between $\eta$ and $\eta'$. Thus, there are $\eta - \eta' \sim \epsilon$ and $1-k \sim \epsilon$. The specific ratio needs to be determined through actual data. In practice, the $\bm{\eta'}$ is a random variable estimated by samples, with mathematical expectation $\eta'$. Thus, the evaluated $\mathbf{k} = k + \bm{\delta}$, where $\bm{\delta}$ is a random variable with 0 center. When the coefficient between $1-k$ and $\epsilon$ is small, reducing the influence of the truncation error. That means that for some data sets, a little $\chi$ is sufficient.

\subsection{Performance}

In Figure~\ref{parameter}, we demonstrate the performance results under different parameter settings. Theoretically, the runtime is decided by the contraction between left environment and $\Gamma$, with complexity $N\chi^2d$. Results in Figure~\ref{parameter} a), b) and c) basically reflects the trend. In Figure~\ref{parameter}, time grows quadratically with $\chi$. When $\chi=10000$, it takes near 2000s to generate 0.4 million samples for Borealis-288 on one A100 GPU, which indicates $\sim$ 14h for 10 milion samples, within acceptable range. That means one can simulate the state-of-the-art photonic quantum computers using \system with only one A100 GPU in less than a day. This is the fastest simulation currently. Figure~\ref{parameter} b) shows a linear but slow growth with $d$. These results implies the influence of other time consumption outside GEMM that does not depend on $d$, which may constitute the next step of optimization space. The variation of time with micro batch $N$ is worth noting, as it is related to memory capacity and arithmetic intensity that we mentioned in our performance model.  That is the key method to meet the extreme data throughput requirements of tensor cores. As the micro batch, $N$ could not be too large, as size of the intermediate tensor is proportional to $N$. If the space for the left environment is occupied, there will be smaller macro batch, leading to worse overlapping and more I/O. However, $N$ should not be too small, since narrow GEMM is inefficient. Figure~\ref{parameter} indicates that the runtime grows slowly at $N < 5000$. Thus, we set $N=5000$, a starting point of linear runtime growth, which ensure enough arithmetic intensity of GEMM to achieve computing bound in Roofline model. In conclusion, these results demonstrates the correctness of our performance models. Both the theoretical and experimental evidences can help to predict performance for a certain optimization and set optimal parameters for larger-scale simulations.

\begin{table}
  \caption{Equivalent bond dimensions of Borealis\cite{madsen2022quantum}, Jiuzhang\cite{zhong2021phase, deng2023gaussian} and a 8176-site simulation data (M8176). The equivalent $\chi$ is calculated as $\sqrt{avg(\chi^2)}$. The "steps ratio" means the proportion of sites that need to be fully calculated. ASP denotes to the number of actual squeezed photons. Parameters are set as: $d=4, \chi=10000$.}
  \label{tab:bond}
    \centering
  \begin{tabular}{ccccc}
    \toprule
    GBS & equi $\chi$ & step ratio & comp ratio & ASP\\
    \midrule
    Jiuzhang2 & 4498 & 0\% & \textbf{20.23\%}& 1.62\\
    Jiuzhang3-h & 7712 & 47.92\% & \textbf{59.47\%} & 3.56\\
    B-M216-h & 8321 & 58.79\% & \textbf{69.23\%} & 6.54\\
    B-M288 & 9132 & 79.51\% & 83.39\% & 10.69 \\
    M8176 & 8923 & 74.29\% & 79.61\% & 8.82\\ 
    \bottomrule
  \end{tabular}
\end{table}

Table~\ref{tab:bond} shows the effect of the dynamic bond dimensions. Our approach achieved a complexity reduction of up to 80\%. The results also shows the positive correlation between equivalent bond dimension and actual squeezed photon number\cite{oh2024classical}. This is in line with the laws of physics, because an increase in the number of actual squeezed photons leads to tighter entanglement.

\begin{figure}[t]
\centerline{\includegraphics[width=0.45\textwidth]{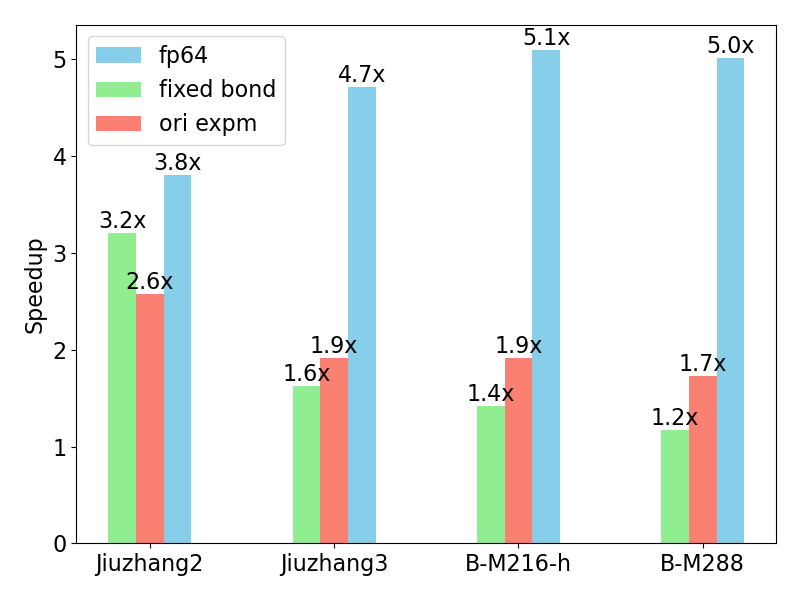}}
\caption{An ablation experiment for our optimization methods on an A100 GPU. Parameters are fixed as $d=4, \chi=10000$, $400K$ samples. In each experiment, our dynamic bond dimension, optimized expm and mixed precision are removed individually to show the impact. The vertical axis represents the speedup of the fully optimized version over the version without certain optimization.}
\label{step}
\end{figure}

Figure~\ref{step} demonstrates the impact of our three main intra-node methods. It is worth noticing that these three methods are decoupled and their optimization effects have a multiplicative effect. So we design an ablation experiment to show their respective effects. As our previous analysis predicts, mixed precision brings huge performance gains on GPU. This implies that our numerically stable algorithm is indispensable for large-scale GBS simulations. Expm optimization stably provides $2\times$ improvements even when $\chi=10000$, which is close to the simulation limit of a single GPU. That means the acceleration will be more significant for smaller $\chi$, which is common in quick verification.

\subsection{Scaling Results}
\begin{figure}[t]
\centerline{\includegraphics[width=0.47\textwidth]{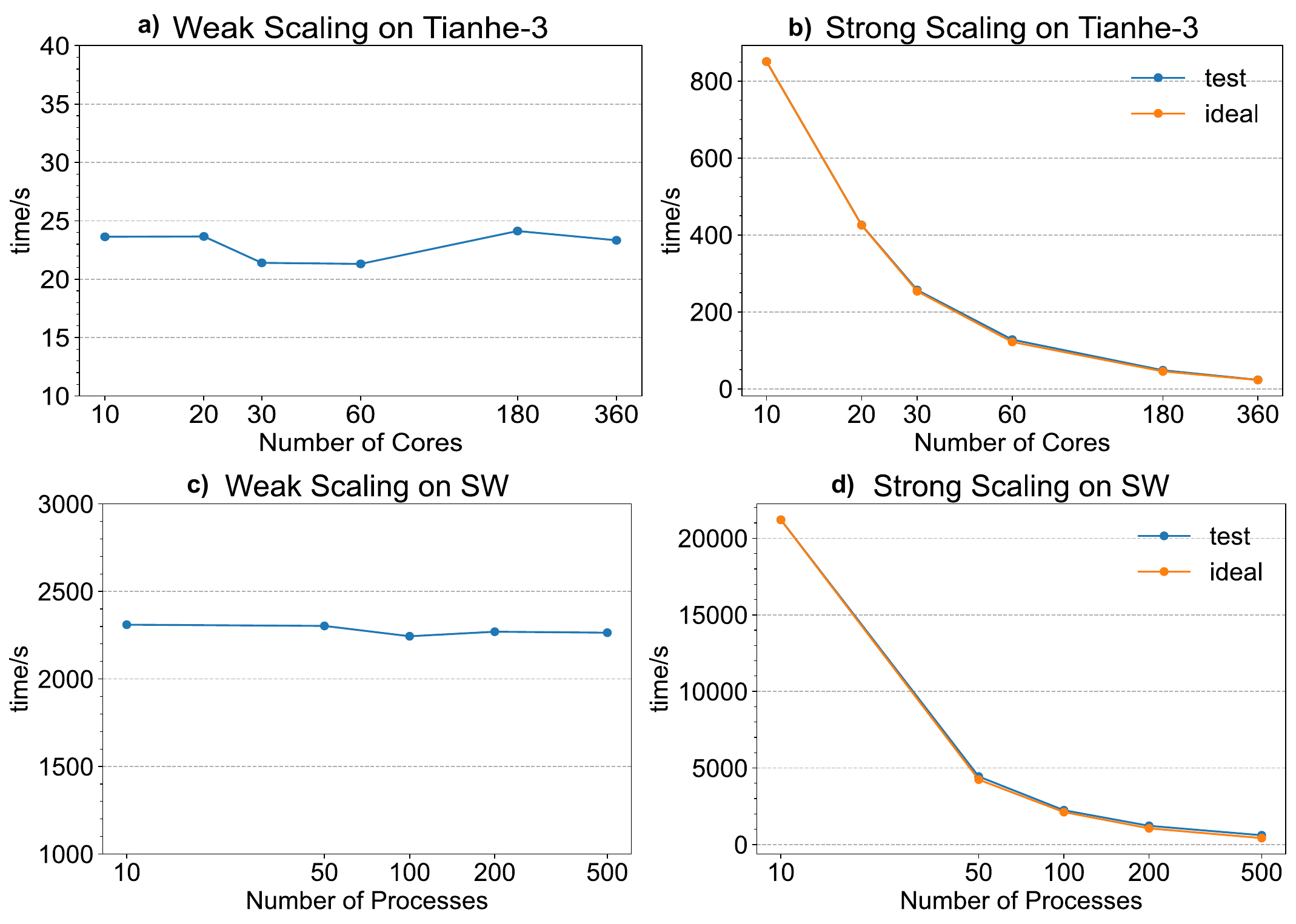}}
\caption{Scaling result of data parallelization on Tianhe-3 and Sunway TaihuLight supercomputer, using the 8176-site simulation data. The parameters are set as: $d=3, \chi=2000$, for Tianhe-3 micro batch size $N_2 = 20000$, and for Sunway $N_2=1000$. Walltime of one site is measured for Tianhe-3 and full 8176-sites for Sunway. a) Weak scaling on Tianhe-3 (1 micro batch each core). b) Strong scaling on Tianhe-3 (360 micro batches). c) Weak scaling on Sunway (5 micro batches each process). d) Strong scaling on Sunway (500 micro batches).}
\label{scaling-th3}
\end{figure}

Since data parallelization occurs between independent samples, and the performance of CPU provides sufficient time to overlap I/O, both of strong scaling and weak scaling are efficient. We tested the scaling results on Tianhe-3 supercomputer and Sunway TaihuLight supercomputer(Figure~\ref{scaling-th3}). On Tianhe-3, due to the resource limitation, we scaled up to 15 nodes and 375 cores and tested one site. On Sunway TaihuLight, we scaled up to 500 processes with 32500 cores with 8176 sites. In weak scaling experiments, each process holds 20000 samples (for Tianhe-3) and 5000 samples (for Sunway). In strong scaling experiments, there are totally 7200000 samples (for Tianhe-3) and 500000 samples (for Sunway). All results achieves more than $95\%$ efficiency, reflecting the low parallel loss and efficient overlapping. 

\begin{figure}[t]
\centerline{\includegraphics[width=0.4\textwidth]{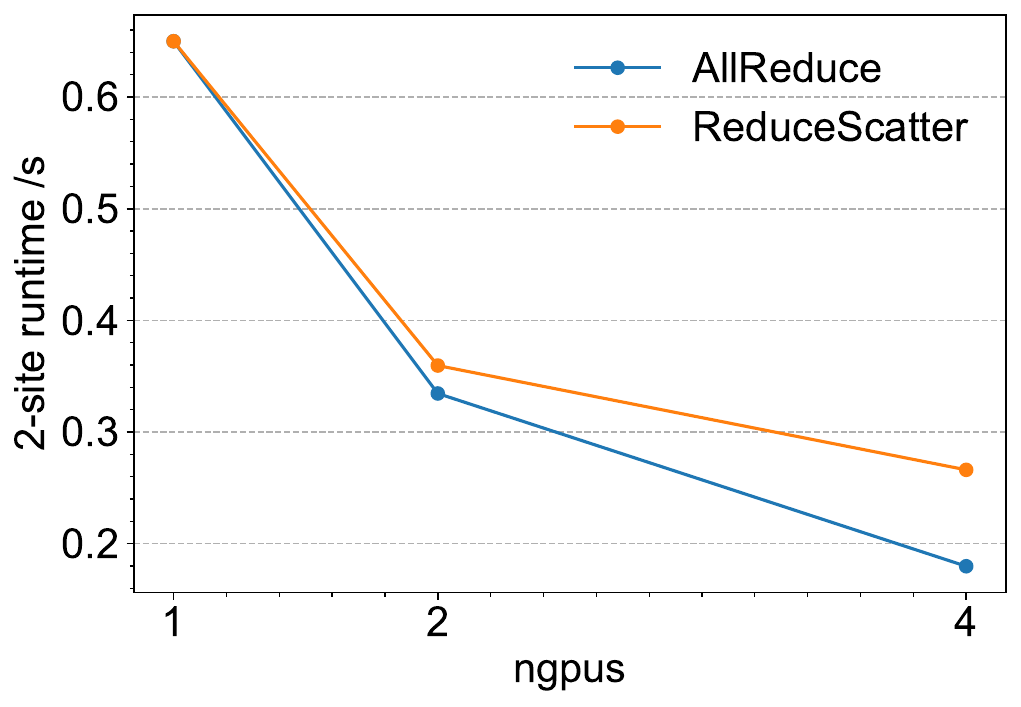}}
\caption{Strong scaling result of tensor parallelization 4 A100 GPUs with 3rd NV-Link, using simulation data: $d=3, \chi=10000, N = 20000$.}
\label{scaling-gpu}
\end{figure}

For tensor parallelization, considering about the collective communication, the situation is slightly different. Here we discuss in two cases, targeting the double-site and single-site parallel scheme respectively. Double-site parallel scheme relies on AllReduce, while single-site parallel scheme relies on ReduceScatter. The scaling result of intra-node tensor parallelization is shown in Figure~\ref{scaling-gpu}. When scaling to 2 GPUs, the communication cost can be ignored. Scaling to 4 GPUs bring $9.8\%$ efficiency decay for double-site scheme and $39\%$ for single-site scheme. For double-site scheme, the overhead mainly comes from measurement time which will be redundantly calculated in the odd sites. There are $T_{Measure} = 0.015s$, $T_{AllReduce}=0.006s$ and $T_{calc} = 0.31s$ in our experiments. While in single-site scheme, despite measurement, the performance is mainly affected by the low bandwidth of ReduceScatter, which accounts for $0.058s$. 

The scaling results above is based on our cluster, with specific hardware and network. For general cases, we designed a simple test as a benchmark, simulating the two communication schemes, to evaluate the performance of AllReduce and ReduceScatter. This benchmark helps to choose scheme on different kinds of clusters.

We use the ratio between communication and computation time to evaluate the overhead of tensor parallelization:
\begin{equation}
    Overhead = (\frac{N_2\chi d}{B} + \eta T_{Measure})/T_{i,N_2}
\end{equation}
There are $\eta=1$ for the double site scheme and $eta=p_2$ for the single site. If the extra overhead is less than $10\%$, we believe that tensor parallelism is effective. Since the communication bandwidth is the key factor of the overhead, GPUs connected with PCIE will be extremely inefficient while doing tensor parallelism. However, with a much lower peak performance, the overhead on CPU will be relatively lower. Our benchmark provides a bandwidth test for AllReduce and ReduceScatter to obtain actual bandwitdh $B_a$ and $B_r$, and also a GEMM profiler to get its peak FLOPS. For example, on four A100 GPU with 3-rd NV-Link connection, there is $B_a = 401GB/s$ and $B_r\sim 46GB/s$, leading to $O_d > O_s$. Thus, we choose double-site scheme to fully utilize the bandwidth. On other machines, decision can be made by testing parameters above.

\subsection{Comparison}

\begin{table}
  \caption{GPU performance results on Borealis\cite{madsen2022quantum} and Jiuzhang\cite{zhong2021phase, deng2023gaussian} experiments (P65-1 for Jiuzhang2), compared with \cite{oh2024classical}. 8 GPUs are organized as $2\times4$ to do data and tensor parallelization. Parameters are set as: $d=4, \chi=10000$, $10M$ samples in total. Time is record in minutes. \system-$n$ refers to $n$ GPUs.}
  \label{tab:comp-gpu}
    \centering
  \begin{tabular}{cccc}
    \toprule
    GBS & MPS\cite{oh2024classical} & \system-1 & \system-8 \\
    \midrule
    Jiuzhang2 & 62 (144 GPUs) & 304.58 & \textbf{38.57}\\
    Jiuzhang3-h & 62 (144 GPUs) & 693.75 & \textbf{95.29} \\
    B-M216-h & 62 (216 GPUs) & 1111.62 & \textbf{152.01}\\
    B-M288 & 62 (288 GPUs) & 1813.75 & \textbf{247.43}\\
    \bottomrule
  \end{tabular}
\end{table}

\begin{table}
  \caption{CPU performance results on Borealis-M288 (B-M288)\cite{madsen2022quantum} and Jiuzhang2\cite{zhong2021phase} experiments, compared with \cite{oh2024classical}. A single core of Intel(R) Xeon(R) Gold 6230R CPU @ 2.10GHz is used for test. Parameters are set as: $d=3, \chi=5000$, $50K$ samples in total.}
  \label{tab:comp-cpu}
    \centering
  \begin{tabular}{cccc}
    \toprule
    GBS & MPS\cite{oh2024classical} & \system & speedup\\
    \midrule
    Jiuzhang2-P65-1 & 17.72h & \textbf{1.76h} & \textbf{10.06}\\
    B-M288 & 36.44h &  \textbf{4.504h} & \textbf{8.09}\\
    \bottomrule
  \end{tabular}
\end{table}

Table~\ref{tab:comp-gpu} and Table~\ref{tab:comp-cpu} illustrate the comparison with \cite{oh2024classical}. We use completely same parameter settings as the GPU experiments in \cite{oh2024classical}. We can generate $10M$ samples for Jiuzhang2 in \textbf{38.57 min} with only 8 GPUs, which outperforms the \textbf{62 min} with 144 GPUs. For Borealis experiments, if assuming a 95\% efficiency with data parallelization (which is supported by our scaling results), \system is estimated to achieve $10.62\times$ and $8.56\times$ speed-up, respectively. The direct comparison with exactly same resource on CPU is in Table~\ref{tab:comp-cpu}. Even without considering higher parallel efficiency, \system can achieve up to $10\times$ speedup on CPU.

\section{Conclusion}

In this work, we introduced \system, a high-performance parallel design to tackle the scalability and efficiency challenges in MPS sampling. We choose photonic quantum computing as the benchmark. Our contributions include a multilevel parallel scheme that ensures scalability and load balancing, an adaptive mixed precision strategy that enables low precision calculation and storage while maintaining numerical stability, and GBS customized optimization. These innovations allow \system to achieve a remarkable 10$\times $ speedup over the existed parallel implementation on A100 GPUs, with the added flexibility of cross-platform portability across x86 CPUs, \emph{Tianhe}, and \emph{Sunway}. 

The ability of \system to simulate large-scale GBS, involving 8176 sites and 16.54 actual squeezed photons, on significantly fewer computational resources represents a substantial advancement in the field. By reducing the dependency on thousands of GPUs and allowing the execution on a single GPU or a small number of nodes, \system makes large-scale GBS simulations more accessible to the research community. These improvements not only keep pace with the rapid progress in photonic quantum computing but also facilitate other applications relying on MPS sampling. 

We find that the parallelization methods of MPS sampling have many similarities with neural networks. If treating the entire MPS as a network, the maximum case in our work will be a large model with 2452B parameters ($M=8176, \chi=10000, d=3$). Therefore, there may be many overlapping with infra in large model inference, with high throughput and high latency. We hope to explore more commonalities between AI and scientific computing applications, so that the development of hardware in the AI-driven era can better promote the acceleration of scientific computing tasks.


\bibliography{main}


\end{document}